# Alternate Timelines for TidalCycles

Alex McLean
Then Try This
Penryn/Sheffield
alex@slab.org

## 1. INTRODUCTION

The TidalCycles (or *Tidal* for short) live coding environment has been developed since around 2009, via several rewrites of its core representation. Rather than having fixed goals, this development has been guided by use, motivated by the open aim to make music. This development process can be seen as a long-form improvisation, with insights into the nature of Tidal gained through the process of writing it, feeding back to guide the next steps of development.

This brings the worrying thought that key insights will have been missed along this development journey, that would otherwise have lead to very different software. Indeed participants at beginners' workshops that I have lead or co-lead have often asked questions without good answers, because they made deficiencies or missing features in the software clear. It is well known that a beginner's mind is able to see much that an expert has become blind to. Running workshops are an excellent way to find new development ideas, but the present paper explores a different technique – the rewrite.

## 2. THE REWRITE

I have re-written Tidal several times before (McLean 2014), or at least largely re-written the inner representation of Tidal patterns and refactored its library of combinators. This involved working in a fresh source folder, but copy-and-pasting a large part of the code, function-by-function from old to new, re-appraising and rewriting as I went. By focussing on the representation, and taking advantages of insights gained since the last rework, generally this involved deleting more code than I wrote. Certainly the type definitions and supporting functions have become significantly shorter and clearer through the process of these rewrites.

However the ongoing rewrite reported on here is as an attempt to recreate Tidal from scratch *without* reference to the existing codebase, as a 'clean room' rewrite. It began with a two-hour public live stream where I 'thought aloud' while rewriting Tidal's core representation of patterns as functions of time. The design of these innards has taken place over 13 years, but this session could be viewed as replaying a condensed version of the thinking behind its design over a mere two hours.

The first motivation for this work was simply curiosity, after participating in a discussion about rewriting generative systems from scratch[1] – how different would it turn out? Would it be better or worse? What insights will be gained, and how will this work feed back into mainline Tidal development? Further motivations have emerged during the rewrite, for example the clarity from refactoring, ideas to formalise Tidal's polymetric sequences underlying its 'mini-notation,' ways to escape Haskell syntax with a custom parser, and facilitating 'ports' of Tidal to other languages.

## 2.1. The First Two Hours

I began the rewrite as a two-hour live stream, to see how much of Tidal I could write in that time.[2] I wasn't sure how interactive or interesting the stream would be, but in the event, I happily talked continuously throughout, with an eye on feedback in the live chat. Surprisingly for me, this felt more engaged than the many times I have streamed musical live coding performances. It felt good to respond to questions and read encouraging messages, while sharing quite an intensive experience of writing the foundations of a system from scratch.

Although this was what you might call "night science," done out of curiosity rather than to respond to a clear research question or design requirement, this approach has some relation to the "think aloud" and "talk aloud" protocols (Ericsson and Simon 1984), which are usability research techniques applied, for example, in the Psychology of Programming field (Wallace et al. 2002) to investigate programming languages as user interfaces. From this experience I have found it possible to 'think aloud' while writing a significant part of a representation for a live coding system, during a relatively short period of time. This points to a potentially useful research programme, where several people are invited to attempt a similar process of thinking aloud while remaking a core part of their system. There has been little in the way of comparative research into the thinking that goes into live coding language design, and this could be a fruitful approach to take. However I only offer this thought for future work, and in-depth analysis of the live stream itself is out of context for this paper.

---

[1] The discussion took place in July 2021, organised by the on-the-fly research group. Mateo Tonatiuh described losing the source code for one of his music systems, remaking it, and finding the new version behaved very differently from the original.

[2] An archive of the live stream is available here: https://youtu.be/F2-evGtBnqQ

The outcome of the stream[3] was 114 code lines[4], roughly two lines per minute, although 37 of these lines were largely redundant type definitions. At the time of writing, additional work tidying and expanding on this work has roughly trebled this line count since, and it is this version that I will compare with the mainline Tidal codebase.[5]

## 2.2. Applicative Functors and Monads

The rewrite focussed on the representation of pattern, and the core functional structures. At this point, Tidal's representation of pattern is well defined, and the rewritten version does not show any significant differences with the original. However the functions representing how patterns are combined look very different.

The rewrite includes the following definition of the standard `<*>` operator and supplemental `<*` and `*>`, which form the basis of Tidal's family of operators for combining patterns (`#`, `+|`, `|*`, and friends). For example these definitions allow two patterns of numbers to be added together, or two patterns of synthesiser effects to be combined, even if the patterns have very different structures. This forms part of Tidal's definition of patterns as an applicative functor. Haskell programmers should note that the use of `<*` and `*>` is unrelated to their usual definition and use in Haskell, rather they define alternate behaviour to `<*>`.

```haskell
app wf patf patv = Pattern f
    where f s = concatMap (\ef -> catMaybes $ map (apply ef) evs) efs
        where efs = (query patf s)
              evs = (query patv s)
              apply ef ev = apply' ef ev (maybeSect (active ef)
                                                   (active ev))
              apply' ef ev Nothing = Nothing
              apply' ef ev (Just s') =
                  Just $ Event (wf (whole ef) (whole ev)) s'
                               (value ef $ value ev)

(<*>) :: Pattern (a -> b) -> Pattern a -> Pattern b
(<*>) = app (liftA2 sect)

(<*) :: Pattern (a -> b) -> Pattern a -> Pattern b
(<*) = app const

(*>) :: Pattern (a -> b) -> Pattern a -> Pattern b
(*>) = app (flip const)
```

---

[3] See the following link for the code resulting from the two hour live stream:
https://github.com/yaxu/remake/commit/8cee36417438e82778b2e0085a2dd897609b8593

[4] Calculated with the *cloc* utility: https://github.com/AlDanial/cloc

[5] The state of the repository at the time of writing:
https://github.com/yaxu/remake/tree/a088f49683f3034881292f20a90d39abc21bdc5f

Unlike in the current 'mainline' Tidal, we can see from the above that the three operators are based on a single function `app`, which does the work of matching events from the pattern of events with the pattern of values. Often, such matching events will only *partly* overlap, are are therefore treated as fragments of an original 'whole' event. This whole is either taken from the pattern of functions (using `<*`), the pattern of values (using `*>`), or the intersection of the two (`<*>`). In the case of combination by addition, the Tidal operators built from these oeprators would be `|+`, `+|` and `|+|` respectively, where structure is said to come from the 'left,' 'right' or 'both.' I won't go into further detail of the workings here, but if I did I would find it a lot easier to write about than the original definitions in 'mainline' Tidal, which are around 2.5 times longer and far less clear.

The following are the rewritten versions of Tidal's 'bind' operations, which again provide core functionality for how patterns are combined. These operations are what makes 'patterned parameters' possible, where for example when using the `fast` function to speed a pattern up, you can also pattern the amount by which it is sped up by. This function (which forms part of the Tidal's definition of a 'monad') is what leads to the popular refrain that in Tidal, it's *patterns all the way down*.

```
bindWhole :: (Maybe Span -> Maybe Span -> Maybe Span) -> Pattern a
          -> (a -> Pattern b) -> Pattern b
bindWhole chooseWhole pv f = Pattern $ \s -> concatMap (match s) $ query pv s
  where match s e = map (withWhole e) $ query (f $ value e) (active e)
        withWhole e e' = e' {whole = chooseWhole (whole e) (whole e')}

bind :: Pattern a -> (a -> Pattern b) -> Pattern b
bind = bindWhole (liftA2 sect)

bindInner :: Pattern a -> (a -> Pattern b) -> Pattern b
bindInner = bindWhole const

bindOuter :: Pattern a -> (a -> Pattern b) -> Pattern b
bindOuter = bindWhole (flip const)
```

Again, this is several times smaller than the current equivalent definitions in 'mainline' Tidal, and far clearer. My reason for including both code snippets here though is to point out the similarities between the definitions of `<*>`, `<*` and `*>`, and those of `bind`, `bindInner` and `bindOuter`. This carries a core insight gained during the rewrite process; when combining two event fragments a common question is, how do you choose what they are a fragment *of*? This insight helps both understand and explain Tidal's approach to combining patterns in time.

I should briefly note that despite the above talk of combining events, Tidal patterns are not data structures, but functions of time. The above functions combine behaviour as functions, but do not actually do any matching of events until the pattern is queried for events, usually by Tidal's scheduler. This is good, because time is infinite, and so procedurally combining all possible events in two patterns would take forever.

## 3. REPRESENTING SEQUENCES

The rewrite has also been an opportunity to rethink how sequences are represented. TidalCycles is often looked upon with suspicion in the Haskell community for its apparent heavy use of strings to represent polymetric sequences as 'mini-notation.' The pejorative term 'stringly typed' (as opposed to 'strongly typed') has been used. To some extent this is a misunderstanding, the double-quoted strings are 'overloaded,' immediately parsed into 'well-typed' functions of time with no other string manipulation. With some exceptions, a rhythm written in the mini-notation is also possible to express directly with Tidal's library of Haskell functions; the mini-notation, inspired by Bernard Bel's *Bol processor* (Bel 2001) simply lets you do so more quickly and succinctly.

Nonetheless, it is true that Tidal does not formally represent the polymetric sequences parsed by the mini-notation. Once they are parsed into a Tidal pattern, the metric structure of the sequence is locked away inside an opaque function of time. This rewrite process was therefore an opportunity to consider how Tidal's current representation of patterns as functions of time could be augmented with equally well formalised representation of rhythms, as polymetric sequences.

To this end, the rewrite currently includes a type for representing rhythm as follows:

```haskell
data Rhythm a = Atom a
  | Silence
  | Subsequence {rSteps   :: [Step a]  }
  | StackCycles {rRhythms :: [Rhythm a]}
  | StackSteps {rPerCycle :: Rational,
                rRhythms  :: [Rhythm a]
               }
  | Patterning {rFunction :: Pattern a -> Pattern a,
                rRhythm   :: Rhythm a
               }
```

An initial aim for this is to be able to represent mini-notation-alike structures in Haskell types, in order to allow Tidal to engage more directly with the world of stateful patterning procedures common in algorithmic music, such as L-Systems, Markov chains, cellular automata and so on. A longer term aim however is to escape Haskell syntax completely. The idea here is that once everything is represented in Haskell, it becomes easier to create something like a mini-notation that embraces the full capability of Tidal. Consider the following piece of design fiction:

```
jux <(rev) (iter 4)> $ sound [bd (every 3 (fast 4) [sn])]
```

The above pattern is not valid Tidal code as we currently know it, and appears to be a jumble of mini-notation and Tidal functions. Indeed, earlier I mentioned that running workshops is a great way to access beginners' minds for fresh perspectives, and I have often seen Tidal beginner workshop participants write code like the above. This raises the question, why *shouldn't* we be able to jumble up these different constructs? We could simply say that `[]` and `{}` allows us to jump into mini-notation-alike rhythms (where multiple subsequences are combined cycle-wise or beat-wise respectively) and `()` allows us to jump back into specifying pattern transformation functions.

In fact, we already have a parser for Tidal as-is that could support such experimentation, the 'MiniTidal' parser created by David Ogborn, used at scale as part of his web-based Estuary live coding environment (Ogborn et al. 2017). There seems to be a path laid out then, for Tidal to first become more clearly formalised in Haskell, in order to then support more flexible, experimental syntax beyond Haskell, such as the above.

## 4. VORTEX - A Tidal 'Port' to the Python Programming Language

This leads to related possibilities offered up by this rewrite – porting Tidal outside of Haskell completely. Now that the core 'innards' of Tidal have been significantly clarified, particularly the applicative and monadic functions shared earlier, it might be easier to port Tidal to other languages. In the past I have sometimes wondered whether it is even practically possible to port Tidal to another language, when its use relies so much on Haskell features of strict types, partial application, type inference and so on. I took the opportunity to find out, by attempting to port Tidal to Python in collaboration with Sylvain Le Beux, Damián Silvani and Raphaël Forment.

It proved to be fairly straightforward to port the core representation and applicative and monadic functions to python, and turning this into something useful became a collective effort, with the fore-mentioned Python programmers in the Tidal community jumping in with ideas and core contributions. The project is codenamed 'vortex,' and around two weeks in, already has a native live coding editor, growing documentation, testing framework, full integration with the Link synchronisation protocol, and is able to make sound via the SuperDirt synthesiser in SuperCollider.

I still doubt that it would have been possible to make something quite like Tidal without the support of Haskell's excellent type system, but now it is made, it seems that the ideas and representations are readily transferable to quite different language paradigms. Nonetheless, there will surely be different trade-offs involved, according to the affordances of the language host, particularly where Tidal takes the form of an embedded domain specific language (as it does with Haskell and now Python).

## 5. CONCLUSION

In summary, this work on rewriting Tidal was driven by curiosity, but lead to unexpected insights and opportunities, in terms of a clearer definitions that opens up new possibilities for experimentation as discussed above.

What I would especially like to highlight in conclusion is the possibility of opening up some of thinking that goes into creating live coding environments. Perhaps we should think about this quite differently from projecting screens in live coding performance, making live coding languages is after all quite a different mode compared to making music, visuals, choreography or other time-based art. Making live coding environments is an opportunity to get very deep in rethinking and restructuring our human relationship with time, where following and properly mapping out an idea might take a decade or more.

Around 13 years into the development of Tidal I feel that I've only recently bee to properly grasp what Tidal is, and am only starting to be able to properly articulate what it is to others. This 'clean room' rewriting process has helped with this aim, and if they haven't already, I encourage other live coding language makers to try doing something similar. This is one way to give ourselves space, as dreamers of ways to dream.

## 6. ACKNOWLEDGMENTS

Special thanks to the Tidal and live coding communities, especially Sylvain Le Beux, Damián Silvani and Raphaël Forment in collaborating on Vortex, the people joining the two-hour Tidal rewrite stream, and the organisers of ,and other participants in, the On-The-Fly research group cantinas which helped keep me thinking during pandemic lockdown.

This research was conducted during the PENELOPE project, with funding from the European Research Council (ERC) under the Horizon 2020 research and innovation programme of the European Union, grant agreement No 682711. It was also conducted during my fellowship as part of Then Try This, supported by a UKRI Future Leaders Fellowship [grant number MR/V025260/1].